\documentclass[useAMS]{iau}
\usepackage{graphicx}
\usepackage{xspace}
\usepackage{amssymb}
\usepackage{comment}
\usepackage{natbib}
\newcommand{\sgra}{Sgr A$^*$\xspace}

\title[X-ray Flares from Sgr A*] 
{The 3 Ms Chandra Campaign on Sgr A*: A Census of X-ray Flaring Activity from the
Galactic Center\vspace{-2mm}}

\author[Neilsen et al.]   
{J.\ Neilsen$^{1,2},$ M.~A.\ Nowak$^{2},$ C.\ Gammie$^{3},$ J.\ Dexter$^{4},$ S.\ Markoff$^{5},$ D.~Haggard$^{6},$  S.\ Nayakshin$^{7},$ Q.~D.\ Wang$^{8},$ N.\ Grosso$^{9},$ D.\ Porquet$^{9},$ J.~A.\ Tomsick$^{10},$ N. Degenaar$^{11},$ P.~C.\ Fragile$^{12},$ J.~C. Houck$^{2},$ R. Wijnands$^{5},$ J.~M.\ Miller$^{11},$ \and F.~K.\ Baganoff$^{2}$}

\affiliation{$^1$Einstein Fellow, Boston University, 725 Commonwealth Ave, Boston MA 02215\\
$^2$MIT Kavli Institute for Astrophysics and Space Research, Cambridge MA 02139\\
$^3$Departments of Astronomy \& Physics, Univ. of Illinois Urbana-Champaign, Urbana IL 61801 \\
$^4$Theoretical Astrophysics Center and Department of Astronomy, UC Berkeley, CA 94720-3411\\
$^5$Astronomical Institute ``Anton Pannekoek'', Univ. Amsterdam, 1098 XH Amsterdam, NL\\
$^6$CIERA Fellow, Physics and Astronomy Department, Northwestern Univ., Evanston, IL 60208\\
$^7$Department of Physics \& Astronomy, University of Leicester, Leicester LE1 7RH UK\\
$^8$Department of Astronomy, University of Massachusetts, Amherst, MA 01002\\
$^9$Observatoire Astronomique de Strasbourg, Universit\'e de Strasbourg, CNRS, 67000 Strasbourg, FR\\
$^{10}$Space Sciences Laboratory, 7 Gauss Way, University of California, Berkeley, CA 94720\\
$^{11}$Department of Astronomy, University of Michigan, Ann Arbor, MI 48109; ND: Hubble Fellow\\
$^{12}$Department of Physics \& Astronomy, College of Charleston, Charleston, SC 29424\vspace{-4mm}
}

\pubyear{2013}
\volume{303}  
\pagerange{119--126}
\jname{The Galactic Center: Feeding and Feedback in a Normal Galactic Nucleus}
\editors{Editors, eds.}
\begin{document}

\maketitle

\begin{abstract}
Over the last decade, X-ray observations of Sgr A* have revealed a black hole in a deep sleep, punctuated roughly once per day by brief flares. The extreme X-ray faintness of this supermassive black hole has been a long-standing puzzle in black hole accretion. To study the accretion processes in the Galactic Center, Chandra (in concert with numerous ground- and space-based observatories) undertook a 3 Ms campaign on Sgr A* in 2012. With its excellent observing cadence, sensitivity, and spectral resolution, this Chandra X-ray Visionary Project (XVP) provides an unprecedented opportunity to study the behavior of the closest supermassive black hole. We present a progress report from our ongoing study of X-ray flares, including the brightest flare ever seen from Sgr A*. Focusing on the statistics of the flares and the quiescent emission, we discuss the physical implications of X-ray variability in the Galactic Center.
\vspace{-3mm}
\end{abstract}\vspace{-4mm}

\firstsection 
\section{Introduction}\label{sec:intro}

In the last fifteen years, X-ray observations of \sgra, the $4\times10^{6}M_{\odot}$ black hole at the center of our Galaxy, have revealed a profoundly quiescent supermassive black hole, its inactivity punctuated roughly once a day by rapid flares (e.g.\ \citealt{Baganoff01,Baganoff03,Dodds-Eden09,Genzel10,Markoff10} and references therein). Short flares are clearly observed from the sub-mm (\citealt{Marrone08}) to the hard X-ray (\citealt{Barriere13}). The flares make a particularly appealing target because, given the paucity of thorough multiwavelength coverage, it has historically been difficult to constrain either the flare mechanism or the dominant radiation process (e.g., synchrotron, synchrotron self-Compton, external Compton).

But despite surpassing the quiescent luminosity of \sgra by factors of 100 or more in the X-ray band (e.g. \citealt{Porquet03,Porquet08,Nowak12}), the flares have actually deepened the puzzle of the ultra-low luminosity of our closest supermassive black hole. Across the mass scale, most weakly accreting black holes fall on the Fundamental Plane of black hole activity (FP), a three-way correlation between black hole mass, X-ray luminosity, and radio luminosity (e.g., \citealt{Merloni03,Falcke04}). \sgra is a notable exception, approaching the FP only during its daily flares (\citealt{Markoff05b,Plotkin12}). These flares therefore trace a link between \sgra and other black holes at very low accretion rates, although these connections cannot explain why \sgra does not \textit{always} fall on the FP. 

\begin{figure*}
\centerline{\includegraphics[width=\textwidth,height=0.49\textwidth]{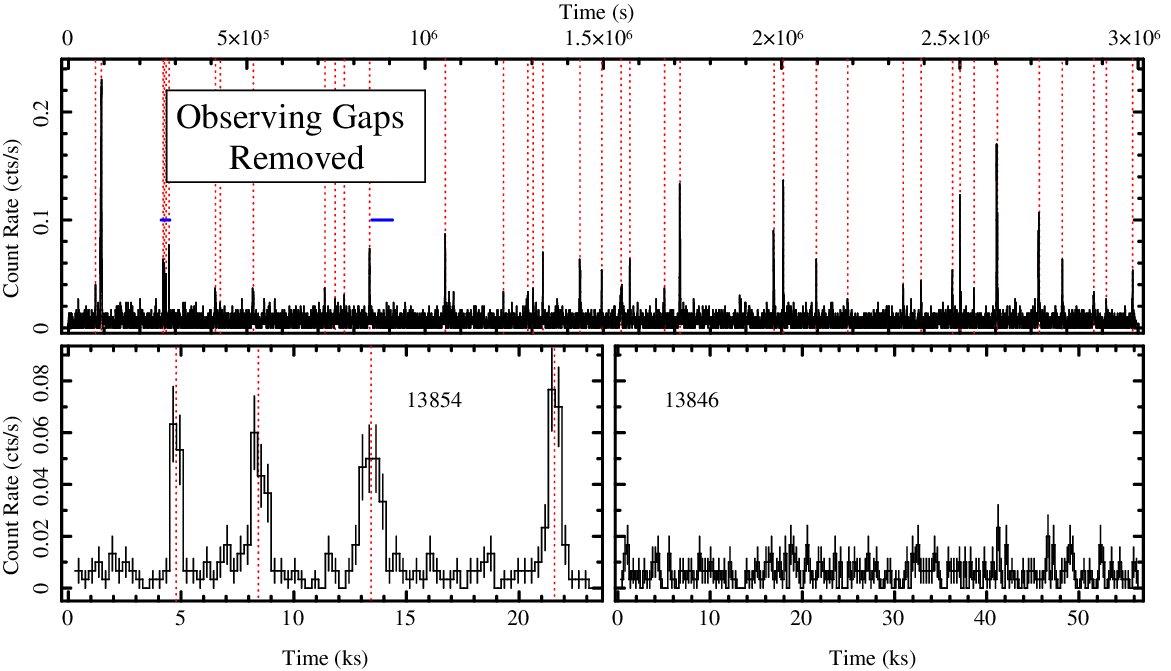}}\vspace{-3mm}
\caption{From \citet{N13b}. (\copyright~2013. The American Astronomical Society. All rights reserved.) (Top): 2--8 keV \textit{Chandra} X-ray lightcurve of \sgra in 300 s bins for the entire 2012 XVP campaign \textit{with gaps removed}. Numerous flares at a range of intensities are apparent, and are identified by dotted red lines. Horizontal lines mark sample observations shown in the bottom plots. (Bottom): Sample light curves of an observation with (left) and without (right) detected flares. ObsID 13854 (left) shows four moderately bright flares within 20 ks. If the flare rate is taken as a constant during 2012, the probability of such a cluster is $\lesssim3.5\%$.\label{fig:lc1}}\vspace{-4mm}
\end{figure*}

The 2012 \textit{Chandra} X-ray Visionary Project (XVP) provides a prime opportunity to study the physics of X-ray flares and their relationship to the quiescent X-ray emission. This 3 Ms campaign to observe the Galactic Center at the highest spatial and spectral resolution available in the X-ray band has the goals of (1) using high-resolution spectra to probe the physics of the accretion flow (see \citealt{Xu06,Young07,Wang13}) and (2) understanding the origin and significance of flares from \sgra. Here, we provide a progress report on our analysis of flares detected during the XVP. Details can be found in \citet{N13b}. \vspace{-6mm}

\section{Flare Properties and Statistics}
Between February and October of 2012, \textit{Chandra} observed the Galactic Center 38 times; the complete light curve of \sgra from the XVP campaign is shown (with observing gaps removed) in Figure \ref{fig:lc1}. A number of bright flares are readily detectable above a steady background (constant within observational uncertainties, this background includes emission from the diffuse gas in the Galactic Center and from the quiescent accretion flow, e.g., \citealt{Wang13}). To detect and characterize these flares, we fit the light curve of each observation with a model consisting of a constant and superimposed Gaussian flares (see \citealt{N13b} for details). We identify 39 flares during the XVP, including the brightest X-ray flare ever observed from \sgra (peak luminosity $L_{\rm X}\sim5\times10^{35}$ erg s$^{-1}$; \citealt{Nowak12}). 

The average observed flare is somewhat more modest, lasting roughly 2600 s and peaking around 0.06 counts s$^{-1}$ (2--8 keV). Scaling from the spectral analysis of the brightest flare, we can estimate a typical mean 2--10 keV flare luminosity of $\sim5\times10^{34}$ erg s$^{-1}.$ The brightest flares seen by \textit{Chandra} and \textit{XMM} have power law spectra with photon indices $\Gamma\sim2$ (\citealt{Nowak12} and references therein), and we see no evidence for any luminosity-dependent variations in the flare hardness ratio (although see \citealt{Degenaar13,Barriere13}). While the analysis of the hardness ratios and spectra of individual flares has not yet conclusively ruled out any flare models, we can glean additional insights from the improved statistics afforded by the XVP.

\begin{figure*}
\centerline{\includegraphics[width=0.5\textwidth,height=0.27\textwidth,trim=0mm 5.75in 0mm 0mm,clip]{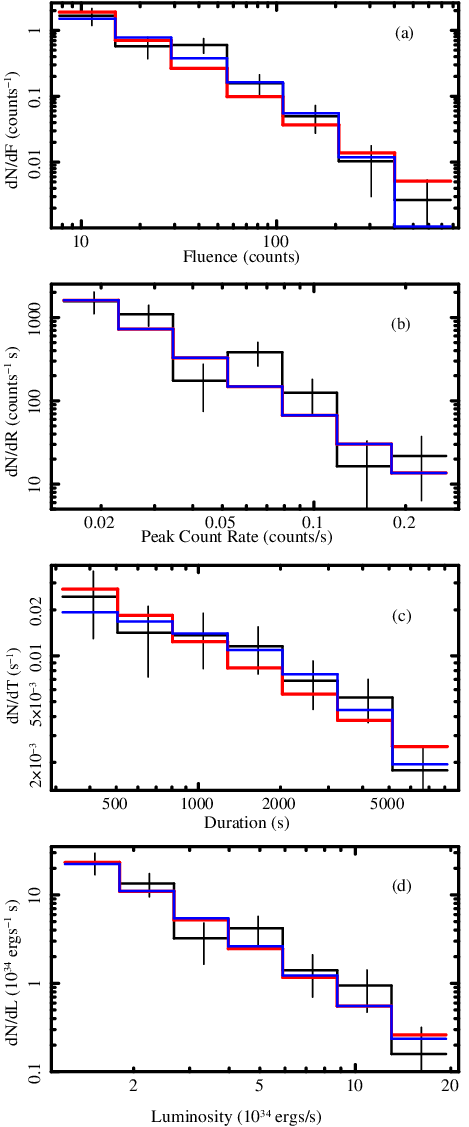}
\includegraphics[width=0.5\textwidth,height=0.28\textwidth,trim=0mm 0in 0mm 5.55in,clip]{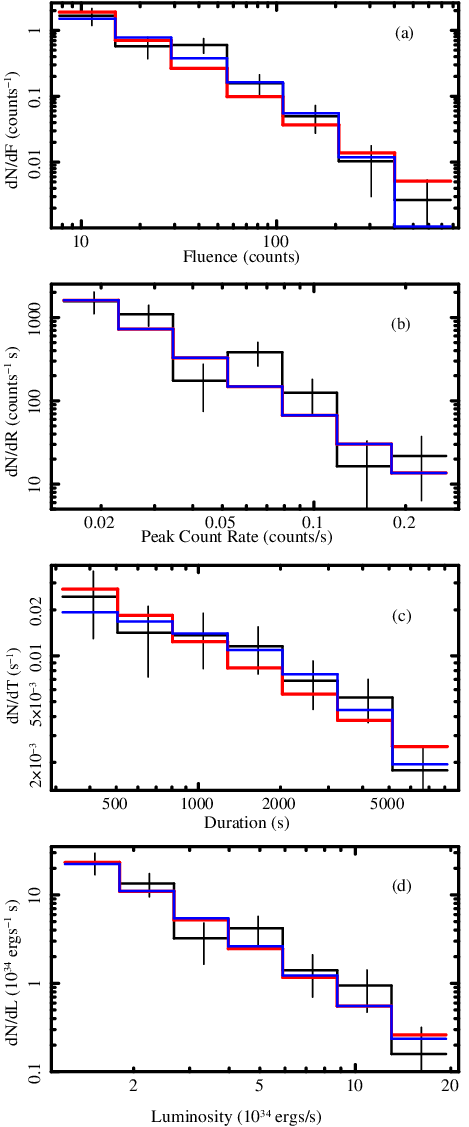}}\vspace{-2mm}
\caption{From \citet{N13b}. (\copyright~2013. The American Astronomical Society. All rights reserved.) Distributions of flare fluence (right) and estimated mean unabsorbed 2--10 keV luminosity (right). The black histograms represent distributions corrected for photon pileup and incompleteness, with Poisson errors on the number of flares in each bin. The red curves are the best power law fit, and the blue curves are the best cutoff power law fit.\vspace{-3mm}\label{fig:hist}}
\end{figure*}

In Figure \ref{fig:hist}, we show the completeness-corrected differential distributions of the fluence (total 2--8 keV counts) and estimated mean 2--10 keV luminosity of the flares. The luminosity distribution is consistent with a power law $dN/dL\propto L^{-1.9\pm0.4},$ which is similar to what is observed in solar flares (\citealt{Crosby11}, although the flares from \sgra are too bright to be stellar in origin) and to what may be expected from the tidal disruption of asteroids if the asteroid mass distribution in the Galactic Center is comparable to that in the solar system (\citealt{Zubovas12}). For other models such as shocks and reconnection (e.g., \citealt{Markoff01}), it is still difficult (if not impossible) to predict luminosity distributions from first principles. 

The fluence distribution scales like $dN/dF\sim F^{-1.5\pm0.2},$ so that the radiant energy is dominated by bright flares. Adding up the observed fluences, we find that 1/3 of the total emitted energy during our 3 Ms campaign was detected in flares, which have a duty cycle of roughly $\sim3.5\%$, so the typical flare is about $\sim10\times$ brighter than the background emission. In addition to counting up the observed flares, we can explore the contribution of undetected flares to the quiescent level. Integrating backwards, below our detection limit, we find that the fluence in undetected flares is $\lesssim10\%$ of the quiescent emission.

\begin{figure}
\centerline{\includegraphics[width=0.5\textwidth,height=0.335\textwidth,trim=0mm 4.22in 0mm 0mm,clip]{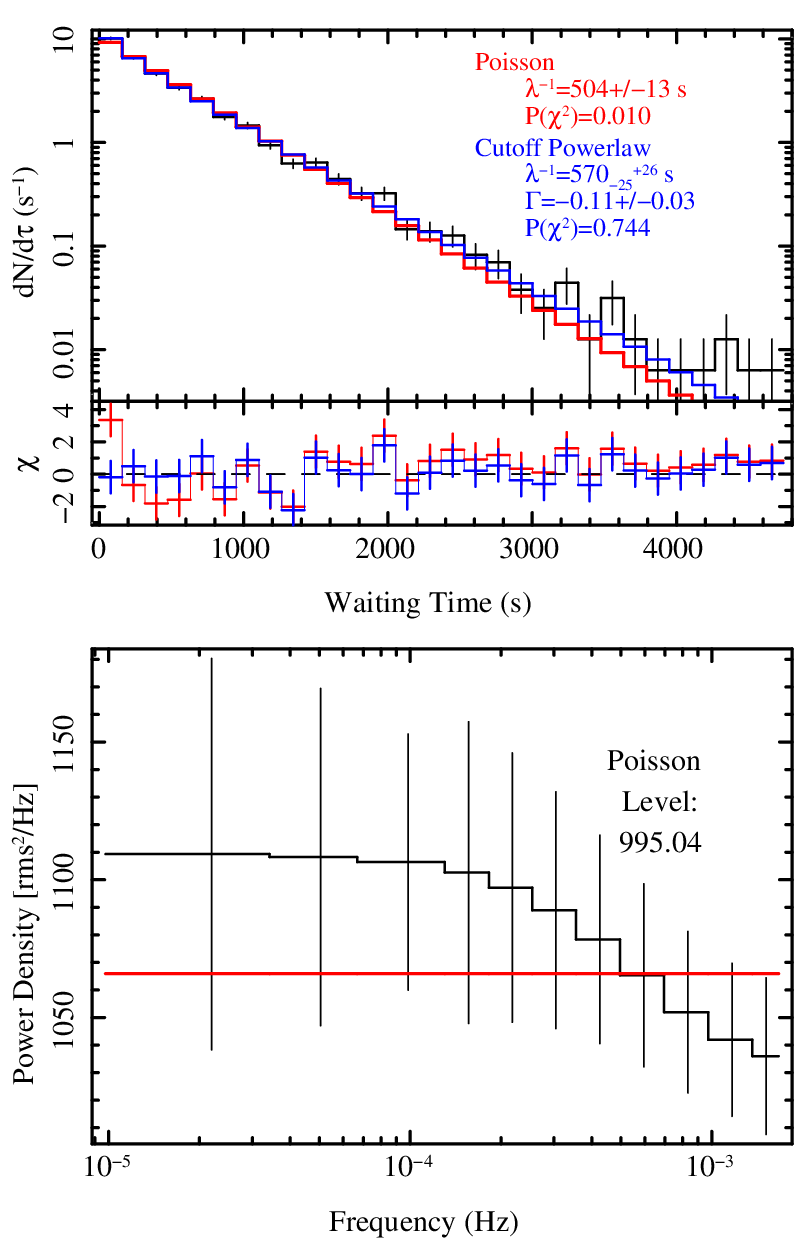}
\includegraphics[width=0.5\textwidth,height=0.32\textwidth,trim=0mm 2mm 0mm 4.25in,clip]{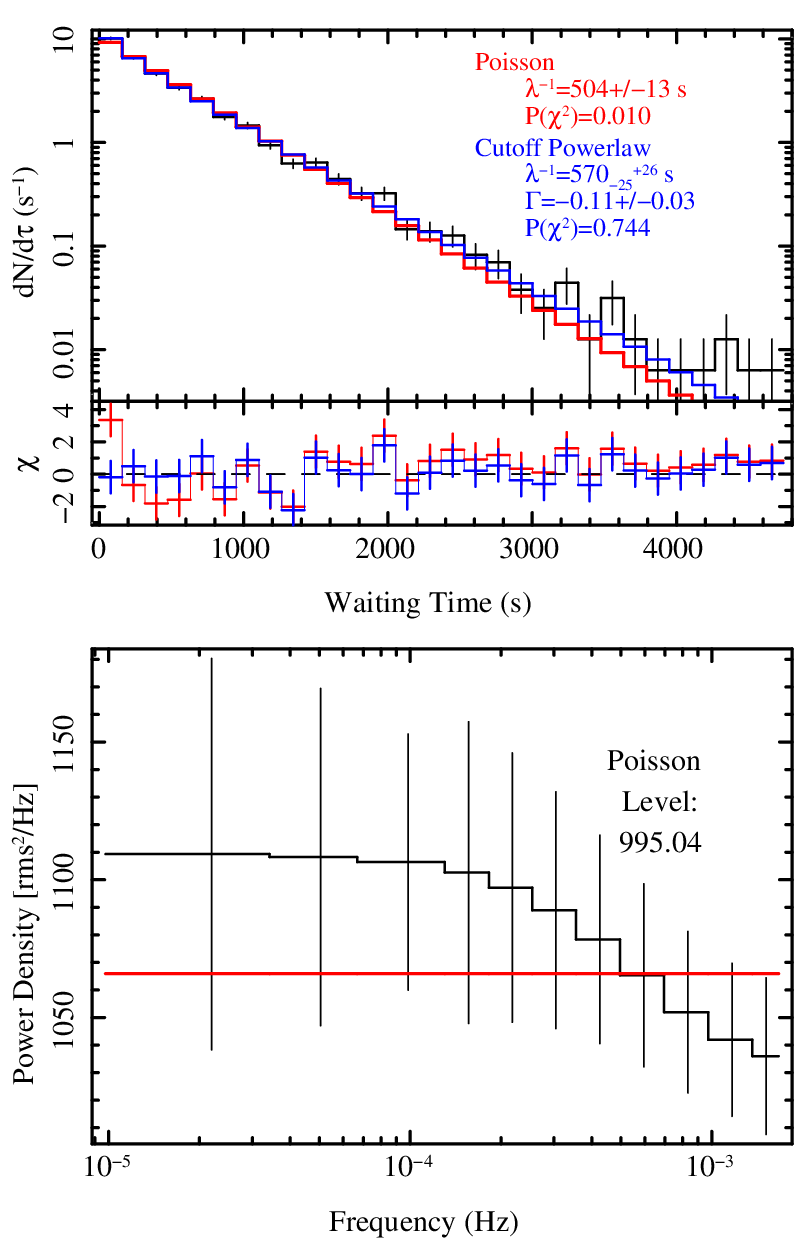}
}\vspace{-3mm}
\caption{From \citet{N13b}. (\copyright~2013. The American Astronomical Society. All rights reserved.) (Left): Distribution of times between quiescent photons, fit with an exponential model and a cutoff power law model (blue). (Right): The estimated power spectrum of the quiescent emission from \sgra is statistically indistinguishable from white noise, and is shown with the best fit constant model. We find a $\sim10\%$ excess above the pure Poisson noise level.\label{fig:poisson}}\vspace{-4mm}
\end{figure}

The contribution of weak/undetected flares to the quiescent emission can also be constrained via variability analysis. In the left panel of Figure \ref{fig:poisson}, we show the observed distribution of waiting times between quiescent photons. For a pure Poisson process (such as might be expected from thermal plasma on scales comparable to the Bondi radius; \citealt{Wang13}), this distribution should be exponential, but we find evidence of an additional component of correlated variability. Monte Carlo estimates of the power density spectrum in quiescence indicate that this component is consistent with white noise, but provides an excess power of $\sim10\%$ over the Poisson level. \vspace{-6mm}

\section{Discussion: Seen and Unseen}
From the \textit{Chandra} light curve in Figure \ref{fig:lc1}, it is clear that the X-ray emission from \sgra comes in at least two flavors: one steady throughout 2012 and one rapidly variable on time scales as short as 100 s (\citealt{Nowak12}). But a close inspection of the quiescent emission reveals that it is not completely steady: there is a $\sim10\%$ variability excess, which could be caused by undetected flares. This interpretation is consistent with a model in which the observed flare distribution extends to very low fluence and undetected flares contribute as much as 10\% of the quiescent flux. Remarkably, spectral analysis of the quiescent emission (\citealt{Wang13}) requires a nonthermal (power-law) component to explain $\sim10-20\%$ of the flux, and models of the surface brightness of \sgra require a point source to explain 10\% of the flux (the rest is extended; \citealt{Shcherbakov10,Wang13}). We conclude that the flare statistics, along with the variability, spectrum, and surface brightness profile of \sgra, provide a consistent physical decomposition of its X-ray emission in quiescence: $\sim90\%$ is steady thermal plasma emission on scales comparable to the Bondi radius, and $\sim10\%$ is power-law emission from flares close to the event horizon. Future statistical studies comparing X-ray and infrared variability (\citealt{Dodds-Eden11,Witzel12}; Neilsen et al., in prep) will provide deeper insight into the radiation physics of these exciting flares.\vspace{-5mm}

\bibliographystyle{apj_set3}
\bibliography{ms}

\begin{thebibliography}{25}
\expandafter\ifx\csname natexlab\endcsname\relax\def\natexlab#1{#1}\fi

\bibitem[{{Baganoff} {et~al.}(2001){Baganoff}, {Bautz}, {Brandt}, {Chartas},
  {Feigelson}, {Garmire}, {Maeda}, {Morris}, {Ricker}, {Townsley}, \&
  {Walter}}]{Baganoff01}
{Baganoff}, F.~K., {et~al.} 2001, \nat, 413, 45

\bibitem[{{Baganoff} {et~al.}(2003){Baganoff}, {Maeda}, {Morris}, {Bautz},
  {Brandt}, {Cui}, {Doty}, {Feigelson}, {Garmire}, {Pravdo}, {Ricker}, \&
  {Townsley}}]{Baganoff03}
---. 2003, \apj, 591, 891

\bibitem[{{Barriere} {et~al.}(2013){Barriere}, {Tomsick}, {Baganoff}, {Boggs},
  {Christensen}, {Craig}, {Grefenstette}, {Hailey}, {Harrison}, {Madsen},
  {Mori}, {Perez}, {Stern}, {Zhang}, {Zhang}, {Zoglauer}, \& {NuSTAR
  Team}}]{Barriere13}
{Barriere}, N., {et~al.} 2013, in AAS/High Energy Astrophysics Division,
  Vol.~13, \#403.02

\bibitem[{{Crosby}(2011)}]{Crosby11}
{Crosby}, N.~B. 2011, Nonlinear Processes in Geophysics, 18, 791

\bibitem[{{Degenaar} {et~al.}(2013){Degenaar}, {Miller}, {Kennea}, {Gehrels},
  {Reynolds}, \& {Wijnands}}]{Degenaar13}
{Degenaar}, N., {et~al.} 2013, \apj, 769, 155

\bibitem[{{Dodds-Eden} {et~al.}(2011){Dodds-Eden}, {Gillessen}, {Fritz},
  {Eisenhauer}, {Trippe}, {Genzel}, {Ott}, {Bartko}, {Pfuhl}, {Bower},
  {Goldwurm}, {Porquet}, {Trap}, \& {Yusef-Zadeh}}]{Dodds-Eden11}
{Dodds-Eden}, K., {et~al.} 2011, \apj, 728, 37

\bibitem[{{Dodds-Eden} {et~al.}(2009){Dodds-Eden}, {Porquet}, {Trap},
  {Quataert}, {Haubois}, {Gillessen}, {Grosso}, {Pantin}, {Falcke}, {Rouan},
  {Genzel}, {Hasinger}, {Goldwurm}, {Yusef-Zadeh}, {Clenet}, {Trippe},
  {Lagage}, {Bartko}, {Eisenhauer}, {Ott}, {Paumard}, {Perrin}, {Yuan},
  {Fritz}, \& {Mascetti}}]{Dodds-Eden09}
---. 2009, \apj, 698, 676

\bibitem[{{Falcke} {et~al.}(2004){Falcke}, {K{\"o}rding}, \&
  {Markoff}}]{Falcke04}
{Falcke}, H., {K{\"o}rding}, E., \& {Markoff}, S. 2004, \aap, 414, 895

\bibitem[{{Genzel} {et~al.}(2010){Genzel}, {Eisenhauer}, \&
  {Gillessen}}]{Genzel10}
{Genzel}, R., {Eisenhauer}, F., \& {Gillessen}, S. 2010, Reviews of Modern
  Physics, 82, 3121

\bibitem[{{Markoff}(2005)}]{Markoff05b}
{Markoff}, S. 2005, \apjl, 618, L103

\bibitem[{{Markoff}(2010)}]{Markoff10}
---. 2010, Proceedings of the National Academy of Science, 107, 7196

\bibitem[{{Markoff} {et~al.}(2001){Markoff}, {Falcke}, {Yuan}, \&
  {Biermann}}]{Markoff01}
{Markoff}, S., {et~al.} 2001, \aap, 379, L13

\bibitem[{{Marrone} {et~al.}(2008){Marrone}, {Baganoff}, {Morris}, {Moran},
  {Ghez}, {Hornstein}, {Dowell}, {Mu{\~n}oz}, {Bautz}, {Ricker}, {Brandt},
  {Garmire}, {Lu}, {Matthews}, {Zhao}, {Rao}, \& {Bower}}]{Marrone08}
{Marrone}, D.~P., {et~al.} 2008, \apj, 682, 373

\bibitem[{{Merloni} {et~al.}(2003){Merloni}, {Heinz}, \& {di
  Matteo}}]{Merloni03}
{Merloni}, A., {Heinz}, S., \& {di Matteo}, T. 2003, \mnras, 345, 1057

\bibitem[{{Neilsen} {et~al.}(2013){Neilsen}, {Nowak}, {Gammie}, {Dexter},
  {Markoff}, {Haggard}, {Nayakshin}, {Wang}, {Grosso}, {Porquet}, {Tomsick},
  {Degenaar}, {Fragile}, {Houck}, {Wijnands}, {Miller}, \& {Baganoff}}]{N13b}
{Neilsen}, J., {et~al.} 2013, \apj, 774, 42

\bibitem[{{Nowak} {et~al.}(2012){Nowak}, {Neilsen}, {Markoff}, {Baganoff},
  {Porquet}, {Grosso}, {Levin}, {Houck}, {Eckart}, {Falcke}, {Ji}, {Miller}, \&
  {Wang}}]{Nowak12}
{Nowak}, M.~A., {et~al.} 2012, \apj, 759, 95

\bibitem[{{Plotkin} {et~al.}(2012){Plotkin}, {Markoff}, {Kelly}, {K{\"o}rding},
  \& {Anderson}}]{Plotkin12}
{Plotkin}, R.~M., {et~al.} 2012, \mnras, 419, 267

\bibitem[{{Porquet} {et~al.}(2008){Porquet}, {Grosso}, {Predehl}, {Hasinger},
  {Yusef-Zadeh}, {Aschenbach}, {Trap}, {Melia}, {Warwick}, {Goldwurm},
  {B{\'e}langer}, {Tanaka}, {Genzel}, {Dodds-Eden}, {Sakano}, \&
  {Ferrando}}]{Porquet08}
{Porquet}, D., {et~al.} 2008, \aap, 488, 549

\bibitem[{{Porquet} {et~al.}(2003){Porquet}, {Predehl}, {Aschenbach}, {Grosso},
  {Goldwurm}, {Goldoni}, {Warwick}, \& {Decourchelle}}]{Porquet03}
---. 2003, \aap, 407, L17

\bibitem[{{Shcherbakov} \& {Baganoff}(2010)}]{Shcherbakov10}
{Shcherbakov}, R.~V., \& {Baganoff}, F.~K. 2010, \apj, 716, 504

\bibitem[{{Wang} {et~al.}(2013){Wang}, {Nowak}, {Markoff}, {Baganoff},
  {Nayakshin}, {Yuan}, {Cuadra}, {Davis}, {Dexter}, {Fabian}, {Grosso},
  {Haggard}, {Houck}, {Ji}, {Li}, {Neilsen}, {Porquet}, {Ripple}, \&
  {Shcherbakov}}]{Wang13}
{Wang}, Q.~D., {et~al.} 2013, Science, 341, 981

\bibitem[{{Witzel} {et~al.}(2012){Witzel}, {Eckart}, {Bremer}, {Zamaninasab},
  {Shahzamanian}, {Valencia-S.}, {Sch{\"o}del}, {Karas}, {Lenzen}, {Marchili},
  {Sabha}, {Garcia-Marin}, {Buchholz}, {Kunneriath}, \&
  {Straubmeier}}]{Witzel12}
{Witzel}, G., {et~al.} 2012, \apjs, 203, 18

\bibitem[{{Xu} {et~al.}(2006){Xu}, {Narayan}, {Quataert}, {Yuan}, \&
  {Baganoff}}]{Xu06}
{Xu}, Y.-D., {et~al.} 2006, \apj, 640, 319

\bibitem[{{Young} {et~al.}(2007){Young}, {Nowak}, {Markoff}, {Marshall}, \&
  {Canizares}}]{Young07}
{Young}, A.~J., {et~al.} 2007, \apj, 669, 830

\bibitem[{{Zubovas} {et~al.}(2012){Zubovas}, {Nayakshin}, \&
  {Markoff}}]{Zubovas12}
{Zubovas}, K., {Nayakshin}, S., \& {Markoff}, S. 2012, \mnras, 421, 1315

\end{thebibliography}

\end{document}